\documentclass[apjl]{emulateapj}
\usepackage{apjfonts,url}
\shorttitle{Nitrogen isotopic ratios in comet 17P/Holmes}
\shortauthors{Bockel\'ee-Morvan et al.}

\begin{document}


\title{Large excess of heavy nitrogen in both hydrogen cyanide and cyanogen from comet 17P/Holmes}


\author{D. Bockel\'ee-Morvan\altaffilmark{1}, N. Biver\altaffilmark{1}, E. Jehin\altaffilmark{2}, A.L. Cochran\altaffilmark{3}, 
H. Wiesemeyer\altaffilmark{4}, J. Manfroid\altaffilmark{2},  
D. Hutsem\'ekers\altaffilmark{2}, C. Arpigny\altaffilmark{2}, J. Boissier\altaffilmark{5}, W. Cochran\altaffilmark{3}, P. Colom\altaffilmark{1}, 
J. Crovisier\altaffilmark{1},  N. Milutinovic\altaffilmark{6}, 
R. Moreno\altaffilmark{1}, J.X. Prochaska\altaffilmark{7}, I. Ramirez\altaffilmark{3}, R. Schulz\altaffilmark{8}, J.-M. Zucconi\altaffilmark{9}}


\email{dominique.bockelee@obspm.fr}

\altaffiltext{1}{LESIA, Observatoire de Paris, 5 Place Jules Janssen, F-92190, Meudon, France}
\altaffiltext{2}{Institut d'Astrophysique et de G\'eophysique, Sart-Tilman, B-4000, Li\`ege, Belgium}
\altaffiltext{3}{Department of Astronomy and McDonald Observatory, University  of Texas at Austin, C-1400, Austin, USA} 
\altaffiltext{4}{IRAM, Avenida Divina Pastora, 7, N\'ucleo Central, E-18012 Granada, Spain}
\altaffiltext{5}{IRAM, 300 rue de la Piscine, Domaine Universitaire, F-38406, Saint Martin d'H\`eres, France}
\altaffiltext{6}{Department of Physics and Astronomy, University of Victoria, 3800 Finnerty Road, Victoria BC V8P 5C2, Canada}
\altaffiltext{7}{Department of Astronomy and Astrophysics, and UCO/Lick Observatory, University of California, 1156 High Street, Santa Cruz, CA 95064, USA}
\altaffiltext{8}{ESA/RSSD, ESTEC, P.O. Box 299, NL-2200 AG Noordwijk, The Netherlands}
\altaffiltext{9}{Observatoire de Besan\c{c}on, F-25010 Besan\c{c}on Cedex, France}



\begin{abstract}
From millimeter and optical observations of the Jupiter-family comet 17P/Holmes performed soon after its huge outburst 
of October 24, 2007, we derive $^{14}$N/$^{15}$N=139$\pm$26 in HCN, and $^{14}$N/$^{15}$N=165$\pm$40 
in CN, establishing that HCN has the same non-terrestrial isotopic composition as CN. The same conclusion 
is obtained for the long-period comet C/1995 O1 (Hale-Bopp) after a reanalysis of previously published measurements. 
These results are compatible with HCN being the prime parent of CN in cometary atmospheres. The $^{15}$N excess  
relative to the Earth atmospheric value indicates that N-bearing volatiles in the solar nebula 
underwent important N isotopic fractionation at some stage of 
Solar System formation. HCN molecules never isotopically equilibrated with the main  
nitrogen reservoir in the solar nebula before being incorporated in Oort-cloud and Kuiper-belt comets. The $^{12}$C/$^{13}$C ratios in HCN and CN are measured to be consistent with the terrestrial value.

\end{abstract}


\section{Introduction}

Comets are made of ices, organics and minerals that record the chemistry of the outer regions of the primitive solar nebula where they agglomerated 4.6 Gyr ago. 
Compositional analyses of comets can provide important clues on the chemical and physical processes that occurred in the early phases of 
Solar System formation, and possibly in the natal molecular cloud that predated the formation of the solar nebula \citep{1}. In particular, isotopic ratios in cometary 
volatiles are important diagnostics of how this matter formed, since isotopic fractionation is very sensitive to chemical and physical conditions. 
 However, such measurements are rare. Strong deuterium enhancements are observed in H$_2$O and 
HCN gases 
of several comet comae that are characteristic of interstellar or protosolar chemistry at low 
temperature \citep{1,2}. On the other hand, questions arise 
of how to explain the non-terrestrial and uniform values of the $^{14}$N/$^{15}$N isotopic ratio in CN \citep{3,4,5,6}, also indicative of interstellar-like chemistry, while 
this ratio was measured to be terrestrial in HCN \citep{8,9}, the presumed source of CN radicals in cometary atmospheres. Here we present 
measurements of the HC$^{14}$N/HC$^{15}$N and H$^{12}$CN/H$^{13}$CN isotopic 
ratios acquired in comet 17P/Holmes by millimeter spectroscopy, together with C$^{14}$N/C$^{15}$N 
and $^{12}$CN/$^{13}$CN measurements from visible spectroscopy. The brightness of this short-period 
(6.9 years) comet of 
the Jupiter family unexpectedly increased (from total visual magnitude $m_v$ = 17 to $m_v$ = 2.5) on October 24, 2007, while it was at a distance of 1.63 AU from the Earth and at 2.44 AU from the Sun (IAU Circ. 8886). This huge outburst of activity, which is   
likely related to a sudden fragmentation of the nucleus followed by the subsequent production of a large quantity of grains, offered us the opportunity to search for weak spectral signatures of rare 
isotopes using complementary techniques.

\section{The $^{14}$N/$^{15}$N ratio in HCN in 17P/Holmes}
\label{sect:2}

\begin{table*}
\caption{Characteristics of radio lines and molecular production rates in comet 17P/Holmes.\label{tbl1}}
\begin{center}
\begin{tabular}{ccccccccc}
\tableline
\tableline
Molecule & Line & Frequency & Date UT          & Int. time &   Line area & Opacity & Column density & Production rate \\
         &      & (GHz)     &  (October 2007)  & (min)     &  (K km s$^{-1}$) &  &(10$^{12}$ mol cm$^{-2}$) & (10$^{26}$ mol s$^{-1}$) \\
\tableline    
HCN     & $J$ = 3--2  & 265.886434 & 27.96--27.98 & 15 &  18.23 $\pm$ 0.14\phantom{0} &   0.56\phantom{0}  & 20.8 $\pm$ 0.17 & 19.85 $\pm$ 0.16\phantom{0}  \\    
HCN     & $J$ = 3--2 $F$ = 2--2& 265.888516 & 27.96--27.98 & 15 & 0.958 $\pm$ 0.083 & 0.03\phantom{0} & 20.7 $\pm$ 1.8\phantom{0}  &  19.74 $\pm$ 1.71\phantom{0}\\    
HCN     & $J$ = 3--2 $F$ = 3--3& 265.884883 & 27.96--27.98 & 15 & 1.097 $\pm$ 0.073 & 0.03\phantom{0} & 23.7 $\pm$ 1.6\phantom{0} & 22.60 $\pm$ 1.50\phantom{0}\\    
H$^{13}$CN   & $J$ = 3--2 &     259.011798 & 28.08--28.19 &  90 & 0.220 $\pm$ 0.037  & 0.008 & 1.74 $\pm$ 0.30 & 0.164 $\pm$ 0.028 \\      
HC$^{15}$N   & $J$ = 3--2 &     258.156996 & 27.96--28.19  &145 & 0.191\ $\pm$ 0.021 & 0.006 & 1.51 $\pm$ 0.16 & 0.140 $\pm$ 0.015 \\      
\tableline
\end{tabular}
\end{center}
\tablecomments{The integration time is on+off source; error bars are 1$\sigma$; line areas refer to the 
main beam brightness temperature scale; column densities and production rates were determined using the  
radiative transfer model of \citet{12} with $x_{ne}$ = 0.5 and a steady state coma with  $T_{kin}$ = 45 K as described in the text. The production rate determined from the whole HCN line 
(resp. HCN hyperfine 
components, H$^{13}$CN and HC$^{15}$N lines) is 6\% lower (resp. 6\% higher), using $T_{kin}$ = 65 K derived by \citet{dello} from HCN 
infrared observations on October 27.6, 2007.}
\end{table*}

\begin{figure}
\epsscale{1.1}
\plotone{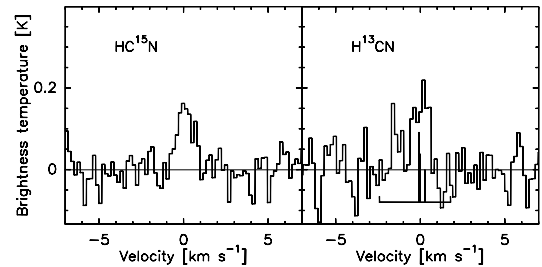}
\caption{Spectra of the $J$ = 3--2 lines of HC$^{15}$N and H$^{13}$CN in comet 17P/Holmes on 
27--28 October 2007. The velocity frame is with respect to the comet rest velocity. 
The positions and relative intensities of the hyperfine components of H$^{13}$CN $J$ = 3--2 are shown.\label{f1}}
\end{figure}

\begin{figure} 
\epsscale{1.1}
\plotone{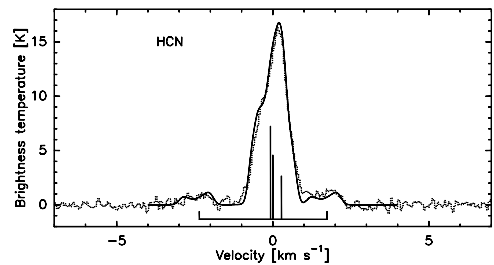}
\caption{Model fitting (continuous line, see text) to the $J$ = 3--2 HCN line profile observed in comet 17P/Holmes on Oct. 27.97 UT (dotted line). Positions and relative intensities of the 
hyperfine components are shown. \label{f2}}
\end{figure}

We carried out observations of 17P/Holmes using the 30-m telescope of the Institut de Radioastronomie Millimétrique (IRAM) located in the Sierra 
Nevada (Spain). Isotopic measurements were performed on 
October 27--28 UT. The tracking of the comet was done using orbital elements K077/06 from JPL Horizons system. 
The pointing of the telescope was checked and updated by repeated observations of a nearby quasar. 
Sky cancellation was 
performed by wobbling the secondary mirror with a throw of 3' at a rate of 0.5 Hz.
Four receivers could be operated at the same time. The $J$ = 3--2 rotational lines of H$^{12}$C$^{14}$N 
(hereafter referred to as HCN) at 265.9 GHz, H$^{13}$CN (259.0 GHz) and HC$^{15}$N  (258.2 GHz) were measured (Table~\ref{tbl1}). The beam diameter (half-power beam width) of 9.5'' corresponded  to 
11300 km at the distance of the comet. Observations were undertaken in good 
atmospheric conditions (3--5 mm precipitable water). The lines were observed both at 
low (1 or 2 MHz) and high (62 kHz) spectral resolutions. Spectra are shown in Figs~\ref{f1}--\ref{f2},
and line areas are given in Table~\ref{tbl1}. Observations of HCN, H$^{13}$CN and 
HC$^{15}$N were not entirely simultaneous (Table~\ref{tbl1}). However, several strong lines (CS $J$ = 3--2, 
CH$_3$CN $J$ = 7--6 (147 GHz) and CH$_3$OH $J$ = 3--2 (145 GHz) lines) were 
continuously observed to monitor the comet activity. Their intensities decreased by 12\% 
during the period (5.5 hours long) when HCN, H$^{13}$CN and HC$^{15}$N data were acquired. This variation was taken into account when deriving the isotopic ratios.

In contrast to the HC$^{15}$N and H$^{13}$CN lines, the HCN $J$ = 3--2 is optically 
thick (Table~\ref{tbl1}). Therefore, radiative transfer modeling 
is required to retrieve the HCN/H$^{13}$CN, HCN/HC$^{15}$N isotopic ratios 
if the HCN $J$ = 3--2 line is used in the analysis. The coma of comet 
Holmes was in a non-equilibrium regime following its outburst, with gas phase 
species released by icy grains.  Unavoidable simplifying model 
assumptions on the coma 
structure could introduce systematic opacity-dependent errors in the retrievals and 
affect the determination of the isotopic ratios. 
However, the HCN $J$ = 3--2 line is split into six hyperfine components, two of which 
($F$ = 2--2, $F$ = 3--3) are well separated from the core of the line, and detected 
in the HCN spectrum (Fig.~\ref{f2}, Table~\ref{tbl1}). These hyperfine components have 
intrinsic line strengths of 3.7\% the total 
strength according to hyperfine statistical weights, and are optically thin 
(Table~\ref{tbl1}). HCN/H$^{13}$CN, HCN/HC$^{15}$N abundance  
ratios determined using the HCN optically thin hyperfine components do not depend 
upon model assumptions on coma temperature, structure and temporal variability, 
because in this case emission lines from molecules in the same excitation state and 
in the same regions of the coma are compared. They are directly given by  
the line intensity ratios, albeit minor corrections accounting for slightly different 
line frequencies. Using the HCN hyperfine components, and correcting for the 
non-simultaneity of the HCN, H$^{13}$CN and HC$^{15}$N  measurements, we derive 
H$^{12}$CN/H$^{13}$CN = 114 $\pm$ 26 and HC$^{14}$N/HC$^{15}$N = 139 $\pm$ 26.  
  
Column densities and production rates determined with the radiative transfer model of \citet{12} are given in  Table~\ref{tbl1}. We assumed a steady-state isotropic 
parent molecule distribution, with a gas velocity of 0.56 km s$^{-1}$, and a gas 
kinetic temperature $T_{kin}$ of 45 K inferred from IRAM 
observations of mutiple lines of CH$_3$OH on October 29.0 UT. 
Interestingly, similar isotopic ratios are obtained using the 
HCN production rate deduced from the whole $J$ = 3--2 line 
as with the model independent hyperfine lines method. 
As shown in Fig.~\ref{f2}, 
when the day and night side HCN velocities are fixed to 0.6 and 0.4  km s$^{-1}$, the 
model provides a satisfactory fit to the shape of the HCN line. This suggests 
that our description of the HCN spatial distribution and excitation is correct in first approximation. This conclusion 
is supported by the good agreement between the HCN production rate measured on October 27.6 in a smaller 
($\sim$ 1'') aperture \citep{dello} and those reported in this work.

\section{The $^{14}$N/$^{15}$N ratio in CN in 17P/Holmes}

\begin{figure}
\begin{center}
\epsscale{.80}
\plotone{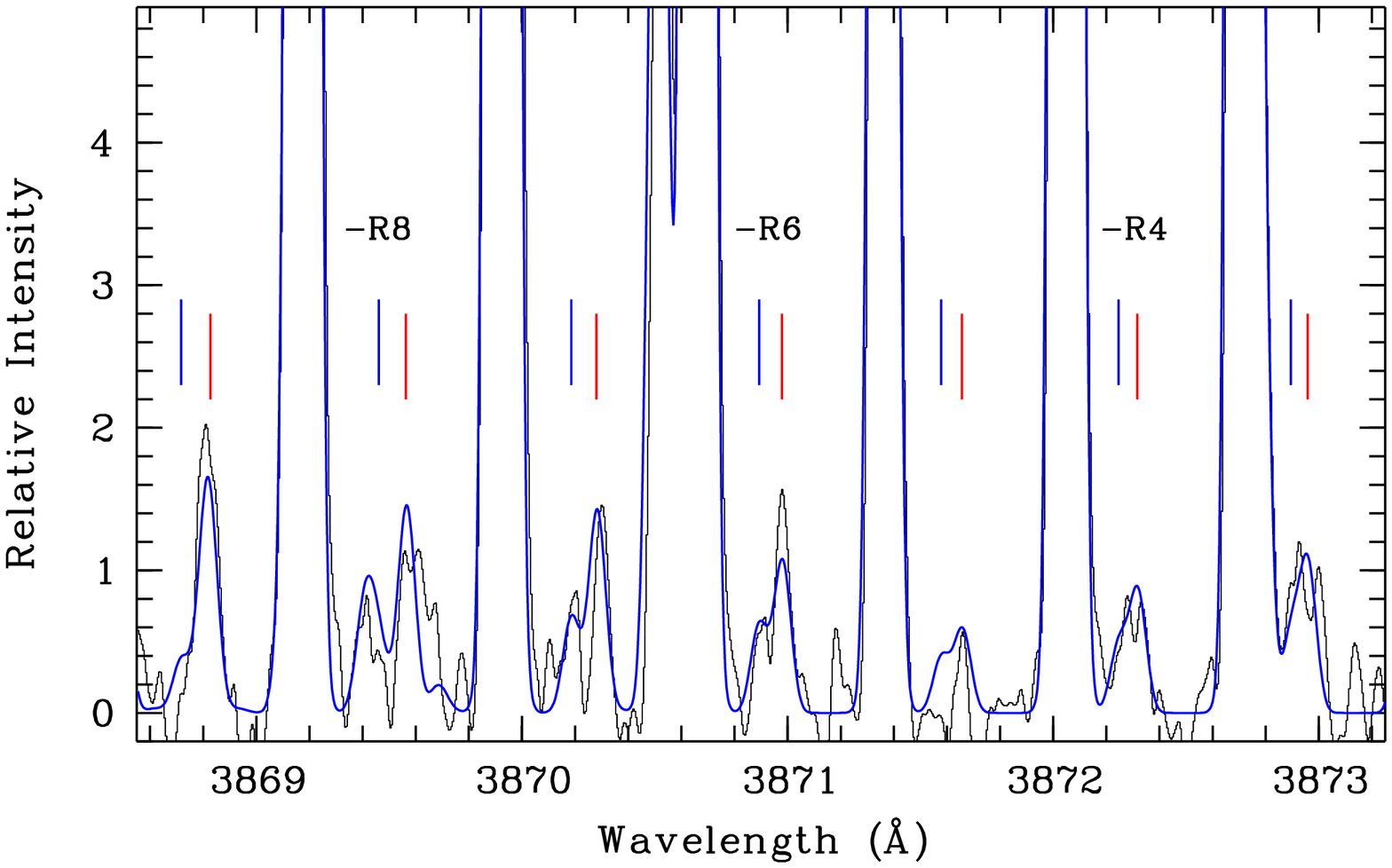}
\end{center}
\caption{
A small part of the co-added Keck and McDonald CN (0,0) band spectra of comet 17P/Holmes showing seven R branch lines (R3 to R9)
 of CN isotopes (black line). The dust-scattered solar spectrum was removed. The positions of the $^{13}$C$^{14}$N (red ticks) and $^{12}$C$^{15}$N 
(blue ticks) lines are indicated; the corresponding lines of the main isotope are on the left. The synthetic spectrum with $^{12}$C/$^{13}$C=90 and $^{14}$N/$^{15}$N = 165 is superimposed (blue line).\label{f3}}
\end{figure}

High-resolution optical observations were performed to 
measure the $^{14}$N/$^{15}$N and $^{12}$C/$^{13}$C ratios in CN. Spectra of the 
B$^2\Sigma^+$--X$^2\Sigma^+$(0,0) CN band at 388 nm were obtained on 
October 25.4, 28.3, 29.4, 30.4, 31.4 and November 18.4 and 19.3 2007 UT 
with the 2DCoudé spectrograph at the 2.7-m Harlan J. Smith telescope of the McDonald 
Observatory. 
A series of short exposures, from 30 seconds to 5 minutes, were also collected 
on October 29.6 2007 UT with the 
High Resolution Echelle Spectrometer (HIRES) of the Keck 1 telescope installed on Mauna Kea (Hawaii). 
The observations were carried out under 
clear weather and low airmass. The slit of the spectrograph was in both cases 
$\sim$1'' wide and $\sim$7'' long providing a resolving power of about 
$\lambda$/$\Delta \lambda$ = 60 000 (0.03 Å/pixel). The slit was centered on the false nucleus in the case of the Keck spectra 
and displaced in the coma (by up to 20'') 
for the McDonald exposures (20 min) to reduce the contamination by the strong dust-reflected spectrum. 
The dust-reflected sunlight underlying the spectral lines (among which are 
ro-vibrational lines of  
 $^{12}$C$^{14}$N, $^{13}$C$^{14}$N and $^{12}$C$^{15}$N (0,0) band) 
was removed by subtracting a solar reference spectrum after the appropriate Doppler shift, 
profile fitting and normalization were applied \citep{3,6}. 
The individual CN (0,0) spectra were then combined with an optimal weighting scheme in order to maximize 
the overall signal-to-noise ratio.  
Synthetic spectra of $^{12}$C$^{14}$N, $^{13}$C$^{14}$N, $^{12}$C$^{15}$N were computed for each observing 
circumstance using a fluorescence model \citep{13}. We took into account 
slightly different excitation conditions 
over the period of the observations caused by small variations of the heliocentric distance and velocity. Collisional effects were empirically estimated by fitting the $^{12}$C$^{14}$N lines \citep{5}. The synthetic spectra were then co-added in the same 
way as the data. The isotope mixture was adjusted to best fit the observed continuum-subtracted 
spectrum. We considered seven R branch lines (R3 to R9), 
as shown in Fig.~\ref{f3}.  The isotopic ratios of $^{12}$C/$^{13}$C and $^{14}$N/$^{15}$N in the final 
co-added Keck and McDonald spectrum are estimated to be 90 $\pm$ 20 and 165 $\pm$ 40, respectively. 

\section{Reanalysis of the comet Hale-Bopp data}

The  $^{14}$N/$^{15}$N ratios measured in HCN and CN in comet 17P/Holmes are consistent 
with each other. In contrast, an HC$^{14}$N/HC$^{15}$N ratio marginally higher than 
the Earth atmospheric value (272) was reported for comet C/1995 O1 (Hale-Bopp) \citep{8,9}, 
to be compared to the value $^{14}$N/$^{15}$N~=~140$\pm$ 35 measured in CN \citep{3}. 
This led us to reanalyse the measurements made in comet Hale-Bopp.

Our reanalysis of the data of~\citet{9} obtained on March 24 and 25, 1997 yields 
$^{12}$C/$^{13}$C = 65 $\pm$ 13 
and $^{14}$N/$^{15}$N = 152 $\pm$ 30 (with a 10\% calibration uncertainty included) as 
compared with the HC$^{14}$N/HC$^{15}$N production rates ratio of 100 $\pm$ 20 and 
286 $\pm$ 82 given by these authors. Ziurys et al. used approximate formulas to 
analyse the optically thick HCN lines, assuming in addition an inappropriate value 
for the rotational temperature of HCN. Instead, we used full radiative transfer 
modeling. Our determinations are likely more reliable as the HCN production rates 
that we derive from the $J$ = 1--0, 2--1 and 3--2 HCN lines observed on March 24 
are consistent within 5\%, while the values inferred by Ziurys et al. from the 
different lines differ by up to a factor of 2. The HC$^{15}$N and H$^{13}$CN 
$J$ = 3--2 lines were observed on March 25. As one cannot exclude day-to-day 
variations in HCN production from March 24 to 25, the HC$^{15}$N/H$^{13}$CN ratio 
of 0.43 $\pm$ 0.10, deduced from March 25 data only, could be more secure. In this case, 
assuming $^{12}$C/$^{13}$C to be equal to the terrestrial value of 89, we 
deduce $^{14}$N/$^{15}$N = 207 $\pm$ 48.
 
The data obtained by \citet{8} at the James Clerk Maxwell 
Telescope (JCMT) are public and available from the Canadian Astronomy Data Centre.
Reanalysing these data, we found: 1) the HCN data were acquired 
during night time near sunrise, 
while the H$^{13}$CN line and especially the HC$^{15}$N line were observed 
later during daytime; the beam efficiency may have degraded by 20\% according 
to JCMT specifications; 2) the HCN line was observed with 
the receiver B3 tuned in double sideband (DSB), while the other lines were observed in 
single sideband (SSB); HCN spectra of calibration sources obtained with the 
same receiver DSB tuning as  
used for the cometary observations show signals in excess of 15\% with respect to reference 
spectra of the sources; 3) the H$^{13}$CN  $J$ = 4--3 line is blended with the 
$\rm SO_2$ $13_{2,12}$--$12_{1,11}$ line at 345.338538 GHz 
\citep{S6}; using the SO$_2$/HCN production rate ratio determined in comet Hale-Bopp \citep{S7}, 
we estimate that it affects the H$^{13}$CN line intensity by 20\%; 
4) more critical, the HC$^{15}$N spectra present scan-to-scan intensity variations by a 
factor of 2 (a factor of 10 above the fluctuations related to the statistical noise) that are likely of instrumental origin. 
Taking these corrections into account, we infer $^{12}$C/$^{13}$C = 94 $\pm$ 8 
and $^{14}$N/$^{15}$N = 205 $\pm$ 70, while the values given in \citet{8} are 
100 $\pm$ 12 and 323 $\pm$ 46, respectively.
The large uncertainty in our $^{14}$N/$^{15}$N determination reflects 
the dispersion of the HC$^{15}$N measurements. 

We conclude that the $^{14}$N/$^{15}$N ratio in HCN is rather uncertain 
for comet Hale-Bopp but is consistent with the value measured in CN. 
  
\section{Implications}

The $^{12}$C/$^{13}$C values in HCN and CN are in agreement with the terrestrial value of 89 and previous measurements in comets 
\citep[e.g.,][]{2}.

The  $^{14}$N/$^{15}$N ratios measured in HCN and CN in comet 17P/Holmes both 
correspond to a factor of two $^{15}$N enrichment relative to the Earth atmospheric value, 
and are consistent with the $^{14}$N/$^{15}$N ratios measured in CN in a dozen comets which cluster at 141 $\pm$ 29 \citep{3,4,5,6}. 
The discrepancy between HCN and CN isotopic ratios previously found for comet 
Hale-Bopp led to the suggestion of a CN production mechanism other than HCN photolysis in cometary atmospheres, possibly from the thermal degradation of $^{15}$N-rich refractory 
organics present in dust grains. This interpretation was supported by the presence of 
CN jets, by the radial distribution of CN, found to be generally less extended than 
expected from HCN photodissociation, and by the CN/HCN production 
rate ratio which exceeds unity in some comets \citep{S8}. However, questions arose 
of how to explain the equally low C$^{14}$N/C$^{15}$N value observed in 
comet Hale-Bopp at large heliocentric distance where the CN radicals were expected to be mainly HCN photodissociation products \citep{5,rauer03}. Our reanalysis of the Hale-Bopp data,  
which shows that the $^{14}$N/$^{15}$N ratio in HCN encompasses the CN value, solves this issue. 
Also, a similar isotopic ratio in HCN and CN provides a better explanation to the uniform values of the C$^{14}$N/C$^{15}$N ratio among comets which exhibit large differences in dust-to-gas production rate ratios. 

Our isotopic measurements are compatible with HCN being the prime parent of CN in cometary 
atmospheres. For comet Hale-Bopp, the production rates of the two species were found 
approximately equal \citep{rauer03,S8}.  
The complex and variable structure of 17P/Holmes's coma after its outburst makes comparisons of the HCN and 
CN production rates difficult for this comet. However, \citet{dello} note that the good 
agreement between the HCN/H$_2$O abundance determined from infrared spectra and the 
OH/CN abundance ratio measured with narrowband photometry \citep{sch07} is consistent with
CN being mainly produced by HCN photolysis. Yet, we cannot exclude that, in other comets, 
CN has other major progenitors (dust or gas-phase species) 
sharing the same low $^{14}$N/$^{15}$N isotopic ratio, which would mean that $^{15}$N enrichment 
is possibly a general property of CN-bearing compounds in comets. 

Comets Hale-Bopp and 17P/Holmes 
are of different dynamical families, the former originating from the Oort cloud, 
and the latter from the trans-Neptunian scattered disc. CN anomalous nitrogen isotopic composition 
is observed in a number of comets of these two dynamical populations. 
Hence, comets issued from these two reservoirs likely exhibit similar 
anomalous N isotopic composition in HCN.

The $^{15}$N excess measured in cometary HCN ($\delta^{15}$N $\sim$ 1000 per mil relative to the Earth 
atmospheric value) and CN 
daughter-product cannot result from isotopic fractionation in the comet atmosphere. It is comparable to the extreme enrichments measured in interplanetary dust particles \citep{14} (IDPs) and carbonaceous meteorites \citep{15}. 
High $\delta^{15}$N values are also present at the submicrometre scale 
in the dust particles collected by the Stardust mission in comet 81P/Wild 2 \citep{16}. In IDPs and meteorites, the $^{15}$N-rich nitrogen is carried by 
non-volatile macromolecular organic material and is generally 
believed to be a remnant of interstellar chemistry, though 
a nucleosynthesis source is often considered as an alternative mechanism 
given the presence of presolar grains with high $^{15}$N excesses in the
matrice of meteorites \citep[e.g.,][]{zin98}. These complex organics 
possibly formed from UV or 
cosmic radiation processing of 
simple ices in the presolar cloud or, at later stages, in the cold regions of the solar nebula. 
The enrichments reported 
here provide the first evidence for the presence of high $^{15}$N anomalies in the volatiles that composed the icy phase of 
the outer solar nebula, possibly representing the precursors of the primitive refractory organics.

Cometary HCN and H$_2$O ices exhibit strong enrichments in deuterium with respect to the cosmic D/H value 
\citep{1,2}, which are believed to reflect ion-molecule and gas-grain fractionation reactions that 
took place at low temperature in the early phases of Solar System formation or in the natal molecular 
cloud. The interpretation of the $^{15}$N enrichment in 
HCN (by a factor of 3 with respect to the protosolar value in the main nitrogen reservoir; 
Fouchet et al. \citeyear{26}; Meibom et al. \citeyear{mei07}) is not as compelling as for deuterium, 
because there is still little evidence for N isotopic 
fractionation in the interstellar medium \citep{23} and predicted $^{15}$N enhancements in HCN for 
exchange reactions involving the main nitrogen reservoir 
N$_2$ are modest \citep{24}. \citet{rod04,rod08} show that highly fractionated NH$_3$ ice could have formed in 
interstellar or protosolar material if N$_2$ was converted into atomic nitrogen. 
While this mechanism is attractive to account for the high $^{15}$N excesses in primitive refractory 
organics, if synthesized from NH$_3$, the chemical 
network for explaining the HCN isotopic anomaly in comets still has to be 
proposed. Alternative mechanisms include a nucleosynthetic origin and
photochemical self-shielding in the solar nebula, similar to that  
proposed for explaining the oxygen isotope anomalies in meteorites 
\citep{clayton02a,clayton02b}. Irrespective of the 
mechanism, protosolar HCN never isotopically 
equilibrated with nebular gas at later phases of Solar System evolution.

\acknowledgments
IRAM is supported by INSU/CNRS (France), MPG (Germany) and IGN (Spain). We thank the IRAM staff, and especially P. Cox for providing us 
discretionary director time, and C. Thum for having scheduled these observations on short notice. Some of the data presented herein were obtained at 
the W.M. Keck Observatory, which is operated as a scientific partnership among the California Institute of Technology, the University of California and the 
National Aeronautics and Space Administration. The Observatory was made possible by the generous financial support of the W.M. Keck Foundation. This paper includes data taken at The McDonald Observatory of The University
of Texas at Austin. We thank G. Punawai for his invaluable assistance during the observing run at Keck 1, and E. Lellouch and D.C. Lis for helpful comments on the manuscript. JM is Research Director, EJ is Research Associate and DH is 
Senior Research Associate at FNRS (Belgium). ALC was funded by NASA Grant 
NNG04G162G.

{\it Facilities:} \facility{IRAM:30m}, \facility{Keck:I}, \facility{McD:2.7m}.



\begin{thebibliography}{}  
\bibitem[Arpigny et al.(2003)]{3}
Arpigny, C., Jehin, E., Manfroid, J., Hutsem{\'e}kers, D., Schulz, R., St{\"u}we, J.~A., 
Zucconi, J.-M., \& Ilyin, I.\ 2003, Science, 301, 1522 

\bibitem[Biver et al.(1999)]{12} 
Biver, N., et al.\ 1999, \aj, 118, 1850

\bibitem[Bockel\'ee-Morvan et al.(2000)]{S7} 
Bockel{\'e}e-Morvan, D., et al.\ 2000, \aap, 353, 1101 

\bibitem[Bockelée-Morvan et al.(2005)]{2} 
Bockelée-Morvan, D., Crovisier, J., Mumma, M. J., \& Weaver, H. A. 2005, in Comets II, ed. M.C. Festou, H.U. Keller, \& H.A. Weaver (Tucson: Univ. of Arizona Press), 391

\bibitem[Busemann et al.(2006)]{15} 
Busemann, H., Young, A.~F., O'D.~Alexander, C.~M., Hoppe, P., Mukhopadhyay, S., \& Nittler, L.~R.\ 2006, Science, 312, 727


\bibitem[Clayton(2002a)]{clayton02a}
Clayton, R.~N.\ 2002a, M\&PSA, 37, A35 

\bibitem[Clayton(2002b)]{clayton02b}
Clayton, R.~N.\ 2002b, Nature, 415, 860 

\bibitem[Dello Russo et al.(2008)]{dello}
Dello Russo, N., Vervack Jr., R.~J., Weaver, H.~A., Montgomery, M.~M., Deshpande, R., Fernandez, Y.~R., 
\& Martin, E.L. \ 2008, \apj, in press

\bibitem[Ehrenfreund et al.(2005)]{1} 
Ehrenfreund, P., Charnley, S.~B., \& Wooden, D.\ 2005, in Comets II, ed. M.C. Festou, H.U. Keller, 
\& H.A. Weaver (Tucson: Univ. of Arizona Press), 115

\bibitem[Floss et al.(2006)]{14} 
Floss, C., Stadermann, F.~J., Bradley, J.~P., Dai, Z.~R., Bajt, S., Graham, G., \& Lea, A.~S.\ 2006, \gca, 70, 2371

\bibitem[Fouchet et al.(2004)]{26} 
 Fouchet, T., Irwin, P.~G.~J., Parrish, P., Calcutt, S.~B., Taylor, F.~W., Nixon, C.~A., \& Owen, T.\ 2004, Icarus, 172, 50 

\bibitem[Fray et al.(2005)]{S8} 
Fray, N., B{\'e}nilan, Y., Cottin, H., Gazeau, M.-C., \& Crovisier, J.\ 2005, \planss, 53, 1243

\bibitem[Hutsemékers et al.(2005)]{4} 
Hutsem{\'e}kers, D., Manfroid, J., Jehin, E., Arpigny, C., Cochran, A., Schulz, R., St{\"u}we, 
J.~A., \& Zucconi, J.-M.\ 2005, \aap, 440, L21 

\bibitem[Ikeda et al.(2002)]{23} 
Ikeda, M., Hirota, T., \& Yamamoto, S.\ 2002, \apj, 575, 250

\bibitem[Jehin et al.(2004)]{6} 
Jehin, E., et al.\ 2004, \apjl, 613, L161 

\bibitem[Jewitt et al.(1997)]{8} 
Jewitt, D., Matthews, H.~E., Owen, T., \& Meier, R.\ 1997, Science, 278, 90

\bibitem[Lis et al.(1997)]{S6} 
 Lis, D.~C., et al.\ 1997, Icarus, 130, 355


\bibitem[Manfroid et al.(2005)]{5} 
Manfroid, J., Jehin, E., Hutsem{\'e}kers, D., Cochran, A., Zucconi, J.-M., Arpigny, C., Schulz, R., \& St{\"u}we, J.~A.\ 2005, \aap, 432, L5 

\bibitem[McKeegan et al.(2006)]{16}
 McKeegan, K. D., et al.\ 2006, Science, 314, 1724
 
\bibitem[Meibom et al.(2007)]{mei07}
Meibom, A., Krot, A.N., Robert, F., Mostefaoui, S., Russel, S., Petaec, M., \& Gounelle, M. \ 2007, \apj, 656, L33


\bibitem[Rauer et al.(2003)]{rauer03} Rauer, H., et al.\ 2003, \aap, 397, 1109


\bibitem[Rodgers \& Charnley(2004)]{rod04} Rodgers, S.~D., \& 
Charnley, S.~B.\ 2004, \mnras, 352, 600

\bibitem[Rodgers \& Charnley(2008)]{rod08} Rodgers, S.~D., \& 
Charnley, S.~B.\ 2008, \mnras, 385, L48

\bibitem[Schleicher(2007)]{sch07}
Schleicher, D.~G. \ 2007, IAUC 8889


\bibitem[Terzieva et al.(2000)]{24} 
Terzieva, R., \& Herbst, E.\ 2000, \mnras, 317, 563


\bibitem[Zinner(1998)]{zin98} 
Zinner, E. 1998, Annu. Rev. Earth Planet. Sci., 26, 147 

\bibitem[Ziurys et al.(1999)]{9} 
Ziurys, L.~M., Savage, C., Brewster, M.~A., Apponi, A.~J., Pesch, T.~C., \& Wyckoff, S.\ 1999, \apjl, 527, L67

\bibitem[Zucconi \& Festou(1985)]{13} 
Zucconi, J.~M., \& Festou, M.~C.\ 1985, \aap, 150, 180

\end{thebibliography}
\end{document}